\documentclass[aps,prl,reprint,superscriptaddress,amsmath,amssymb,amsthm,mathrsfs]{revtex4-1}

\usepackage{graphicx}
\usepackage{dcolumn}
\usepackage{bm}
\usepackage[ 
margin=0.75in,
]{geometry}

\def\TN{T$_\mathrm{N}$}
\def\abinitio{\textit{ab initio}}
\def\dr{$d(r)$}
\def\gr{$G(r)$}
\DeclareMathAlphabet{\mathpzc}{OT1}{pzc}{m}{it}

\begin{document}
	
	\title{Verification of Anderson superexchange in MnO via magnetic pair distribution function analysis and \textit{ab initio} theory}
	
	\author{Benjamin A. Frandsen}
	\affiliation{Department of Physics, Columbia University, New York, NY 10027}
	
	\author{Michela Brunelli}
	\affiliation{Swiss Norwegian Beamlines, European Synchrotron Radiation Facility (ESRF), Grenoble, France}
	
	\author{Katharine Page}
	\affiliation{Spallation Neutron Source, Oak Ridge National Laboratory}
	
	\author{Yasutomo J. Uemura}
	\affiliation{Department of Physics, Columbia University, New York, NY 10027}
	
	\author{Julie B. Staunton}
	\affiliation{Department of Physics, University of Warwick, Coventry CV4 7AL, United Kingdom}

	\author{Simon J. L. Billinge}
	\email{sb2896@columbia.edu}
	\affiliation{%
		Condensed Matter Physics and Materials Science Department, Brookhaven National Laboratory, Upton, NY 11973, USA.
	}%
	\affiliation{%
		Department of Applied Physics and Applied Mathematics, Columbia University, New York, NY 10027, USA.
	}%
	
	\date{\today}
	
	\begin{abstract}
		We present temperature-dependent atomic and magnetic pair distribution function (PDF) analysis of neutron total scattering measurements of antiferromagnetic MnO, an archetypal strongly correlated transition-metal oxide. The known antiferromagnetic ground-state structure fits the low-temperature data closely with refined parameters that agree with conventional techniques, confirming the reliability of the newly developed magnetic PDF method. The measurements performed in the paramagnetic phase reveal significant short-range magnetic correlations on a $\sim$1~nm length scale that differ substantially from the low-temperature long-range spin arrangement. \textit{Ab initio} calculations using a self-interaction-corrected local spin density approximation of density functional theory predict magnetic interactions dominated by Anderson superexchange and reproduce the measured short-range magnetic correlations to a high degree of accuracy. Further calculations simulating an additional contribution from a direct exchange interaction show much worse agreement with the data. The Anderson superexchange model for MnO is thus verified by experiment and confirmed by \abinitio\ theory.
	\end{abstract}
	
	
	\maketitle
	
	Magnetic monoxide MnO, because of its relative simplicity, is often used as a benchmark system for the development of theoretical methods for understanding properties of transition metal oxides~\cite{ander;pr59}.  For example, various tight-binding models~\cite{takah;prb96,harri;prb07,harri;prb08} and many important implementations of density functional theory (DFT)~\cite{pask;prb01,fisch;prb09,tran;prl09,thuns;prl12,sakum;prb13,das;prb15} that include the effects of strong electron correlations have been tested against MnO.
	Modern applications of these \abinitio\ methods can successfully determine gross features of MnO, including the average atomic structure, magnetic ordering temperature, and local magnetic moment size. Unfortunately, efforts to draw conclusions about the underlying physics from these computations are frequently complicated by the fact that differing theories often rely on significantly different assumptions and make widely varying predictions about important details of the system. Among others, these details include the precise orientation of the magnetic moments in the antiferromagnetic (AF) phase, the nature of the short-range magnetic correlations above \TN, and the mechanism(s) for magnetic exchange.
	
	Experimental techniques capable of probing local structure and properties can address these issues in ways not possible with bulk probes of average properties. Illustrating this are recent neutron total scattering experiments~\cite{goodw;prl06,frand;aca15} that revealed a previously hidden local monoclinic distortion in MnO at low temperature and a preferred spin orientation axis along [10$\bar{1}$] or [11$\bar{2}$]. This new information, inaccessible by standard characterization techniques  such as magnetic susceptibility and conventional neutron diffraction, can then be used to differentiate between competing theories.  In this Letter, we utilize magnetic pair distribution function (mPDF) analysis~\cite{frand;aca14}, a recently developed method for investigating local magnetic structure, to measure directly the short-range magnetic correlations in the paramagnetic state of MnO from temperature-dependent neutron total scattering experiments. We use these results to evaluate competing theories of magnetic exchange in MnO, finding that the Anderson superexchange~\cite{ander;pr59} obtained from recent DFT calculations with the self-interaction-corrected (SIC) local spin density approximation~\cite{perde;prb81,luder;prb05} in the ``disordered local moment'' (DLM) approach~\cite{gyorf;jppfm85,hughe;njp08} describes the data exceptionally well with no need for an additional direct exchange contribution present in other models. In addition to resolving this longstanding question about MnO, this work highlights the mPDF technique and the DLM-DFT(+SIC) scheme as valuable tools for studying magnetic properties of strongly correlated electron systems.	

	Neutron time-of-flight total scattering measurements were performed at the NPDF instrument at the spallation source located at the Lujan Neutron Scattering Center at Los Alamos National Laboratory on a commercial sample of MnO (Alfa Aesar, 99\% pure). Low backgrounds and wide momentum-transfer coverage contributed to high-quality data appropriate for magnetic and atomic pair distribution function (PDF) analysis. The data were reduced for PDF transformation according to standard protocols within the program PDFgetN~\cite{peter;jac00}. A closed cycle refrigerator was used to access temperatures between 15~K and 300~K. We also conducted neutron scattering measurements on the same sample at the D20 instrument of the Institute Laue-Langevin (ILL) reactor source, using a neutron wavelength of 0.94079~\AA\ and an Orange cryostat for temperatures in the range $5< T<300$~K.
	
	Similar to the atomic pair distribution function (PDF)~\cite{egami;b;utbp12}, the mPDF technique involves normalizing and Fourier transforming the magnetic scattering intensity from a powder sample to obtain the pairwise magnetic correlation function directly in real space. Because both Bragg and diffuse scattering are included, the mPDF is sensitive to both long- and short-range magnetic correlations. Simplistically speaking, pairs of spins with parallel alignment contribute a positive peak in the mPDF and anti-parallel spins give a negative peak, providing a highly intuitive view of spin correlations on a local length scale.  The full definition of the mPDF equations are in Ref.~[\onlinecite{frand;aca14}]. 
	
	In practice, the mPDF is most easily obtained when it is done simultaneously with the atomic PDF \gr~\cite{frand;aca15}, without separating the magnetic and nuclear scattering signals or normalizing the magnetic scattering by the magnetic form factor, $f_m(Q)$ as originally proposed~\cite{frand;aca14}. The unnormalized mPDF quantity \dr\ obtained this way is given by~\cite{frand;aca15}
	\begin{align}
		d(r)=C_1 \times \mathpzc{f}(r)\ast S(r) + C_2 \times \frac{\textrm{d}S}{\textrm{d}r},\label{eq;dr}
	\end{align}
	where $f(r)$ is the properly normalized mPDF derived in Ref.~[\onlinecite{frand;aca14}], $C_1$ and $C_2$ are constants related by $C_1 / C_2 = -1 / \sqrt{2\pi}$ when the ordered moment is fully saturated (the ratio will be smaller in magnitude otherwise), $\ast$ represents the convolution operation, and $S(r)=\mathcal{F}\left\{f_{m}(Q)\right\}\ast \mathcal{F}\left\{ f_{m}(Q)\right\}$, where $\mathcal{F}\left\{\Box\right\}$ denotes the Fourier transform. The second term in this equation, which appears as a peak at very low $r$, contains no information about magnetic correlations and is determined entirely by the spatial distribution of the localized Mn$^{2+}$ magnetic moment. This equation provides a straightforward method to calculate \dr\ from a magnetic structure model for direct comparison with the experimental mPDF and has been implemented as an extension to the Diffpy-CMI software package~\cite{juhas;aca15}, and this approach to modeling the mPDF was used in the current work.
	
	\begin{figure}
		\includegraphics[width=80mm]{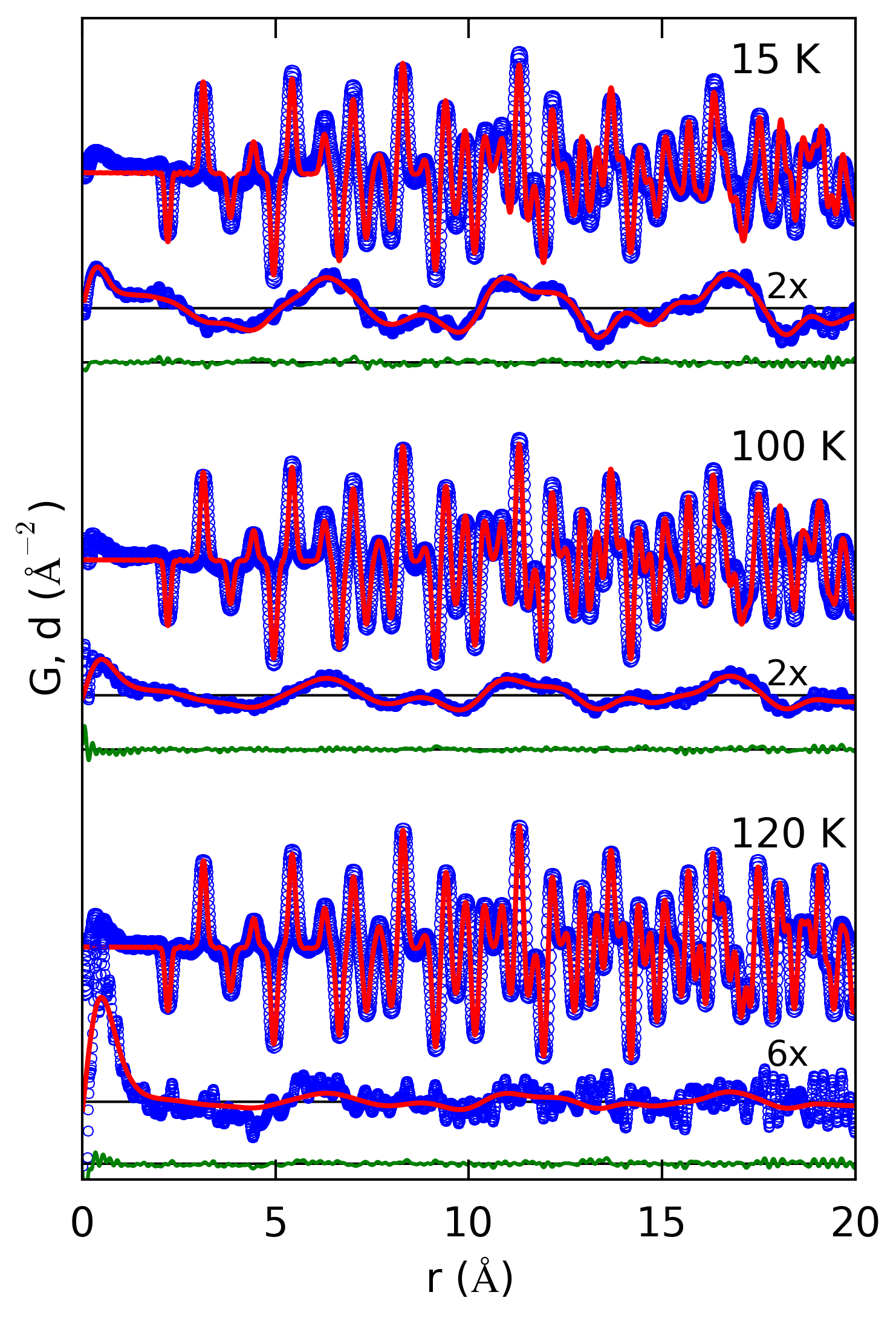}
		\caption{\label{fig:LRO} (Color online) Combined atomic and magnetic PDFs of MnO at representative temperatures. For each temperature, the combined atomic and magnetic PDF is shown as the top blue curve, with the refined atomic PDF overlaid in red; the residual from the atomic PDF fit is shown as the lower blue curve, with the refined mPDF overlaid in red (both magnified by the factor indicated for clarity); and the green curve gives the overall fit residual.}
	\end{figure}
	
	Fig.~\ref{fig:LRO} displays the combined atomic and magnetic PDF obtained from the NPDF instrument at three representative temperatures: 15~K (deep in the ordered state), 100~K, and 120~K (just above the AF transition at \TN$= 118$~K). For each temperature, we first refined the rhombohedral structural model and then used the residual after subtraction of the calculated \gr\ to refine the magnetic structure, modeled by the known long-range Type II antiferromagnetic structure and constrained by the atomic positions from the structural fit. This magnetic structure consists of ferromagnetically aligned spins within common (111) sheets and antiferromagnetic coupling between adjacent sheets. All fits were performed from 0~\AA\ to 20~\AA. For each temperature, the top set of curves shows in blue the total experimental PDF (structural and magnetic) and in red the refined structure-only PDF \gr; the second set shows in blue the residual signal after subtraction of the calculated \gr\ and in red the refined mPDF \dr\ (magnified for clarity as indicated in the figure); and the lower green curve shows the combined PDF+mPDF fit residual, revealing an excellent overall fit quality. Although previous work~\cite{goodw;prl06,frand;aca15} demonstrated that the local structure at low temperature has a slight monoclinic distortion, we use the rhombohedral model at all temperatures for simplicity. Interestingly, our structural refinements with the rhombohedral model indicate that a slight rhombohedral distortion persists on a $\sim 5-10$~nm length scale in the high-temperature cubic phase. Details of this local distortion are provided in the Supplementary Information. 
	
	The mPDF signal can be easily distinguished from the atomic PDF by two features: it is lower in frequency and characteristically broader than the atomic PDF due to the effects of the magnetic form factor, and it changes more dramatically with temperature. This allows robust refinement of the atomic and magnetic structures with no risk of confounding the fits. We noticed that the mPDF peak at low $r$ arising from the second term in Eq.~\ref{eq;dr} is rather independent of temperature, as expected when the local moment size and spatial distribution are constant, so we performed a global fit to this feature at all temperatures to determine $C_2$, which we fixed for subsequent analysis. Imposing the known long-range AF structure of MnO left the scale factor $C_1$ as the only free parameter in our mPDF fits. This scale factor is proportional to the square of the ordered magnetic moment, similar to the intensity of a magnetic Bragg peak for long-range magnetic order.
	\begin{figure}
		\includegraphics[width=80mm]{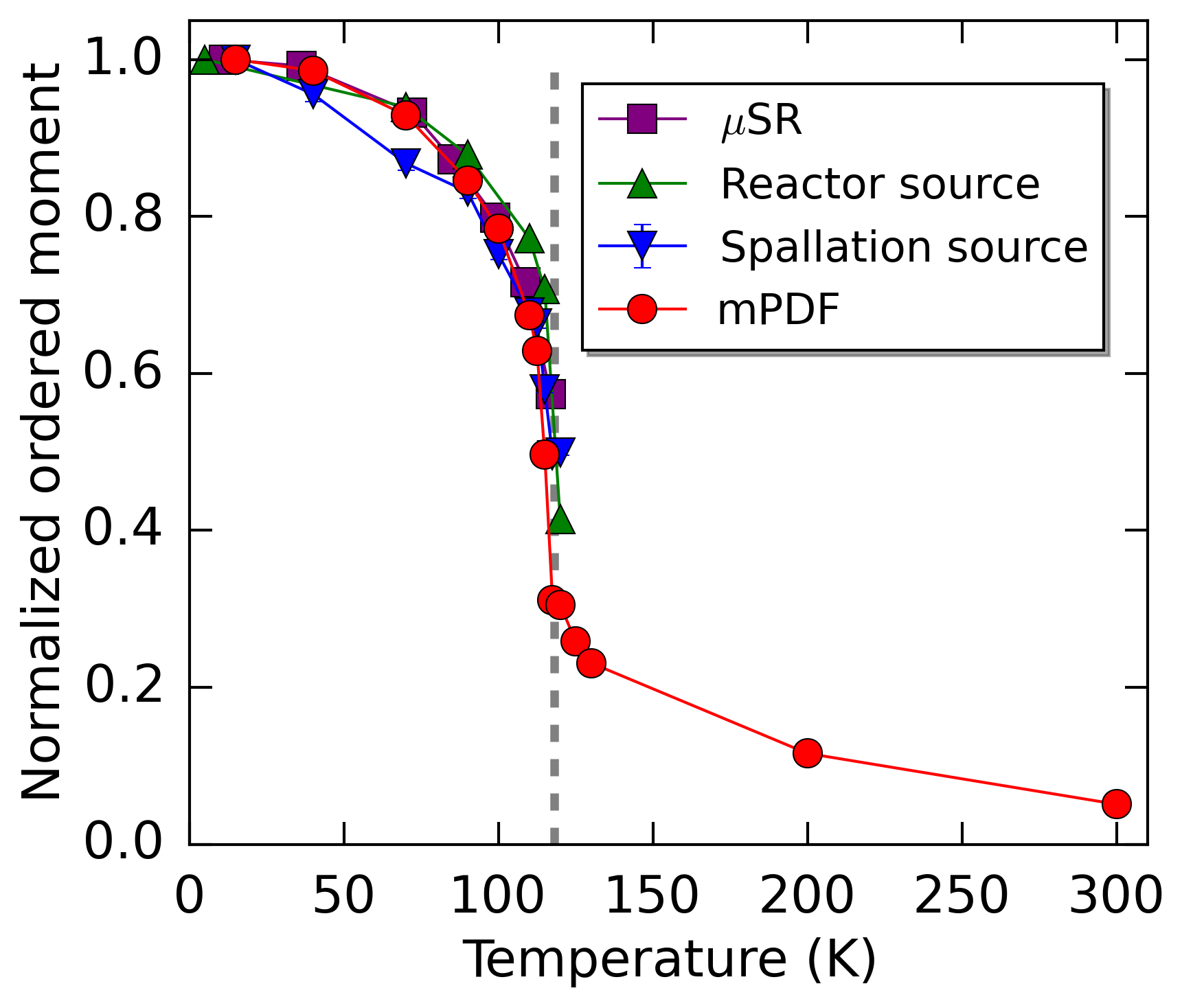}
		\caption{\label{fig:OP} (Color online) Magnetic order parameter obtained from different experimental techniques, including the $\mu$SR precession frequency, the maximum AF Bragg peak intensity from a reactor source, the integrated AF Bragg peak intensity from a spallation source, and the refined mPDF fit of the known magnetic structure of MnO at low temperature. The vertical dashed line indicates the magnetic ordering temperature of 118~K.}
	\end{figure}
	In Fig.~\ref{fig:OP}, we plot $\sqrt{C_1}$ as a function of temperature (normalized by its maximum value) as red dots and compare it to the normalized magnetic order parameter determined by the precession frequency in a muon spin rotation ($\mu$SR) experiment~\cite{uemur;hfi84}, the square root of the maximum intensity of the first AF Bragg peak obtained from our ILL reactor source measurements, and the square root of the integrated intensity of the same AF Bragg peak from the total scattering structure function $S(Q)$ of the time-of-flight neutron data. The results from these approaches mutually confirm each other by their close agreement below \TN\ and demonstrate the reliability of the mPDF approach for analyzing magnetic scattering data.
	
	More interesting is the behavior of the mPDF scale factor above \TN: we observe that $\sqrt{C_1}$ remains finite but decays approximately exponentially with temperature above \TN, confirming the presence of short-range magnetic correlations in the paramagnetic state. This in itself provides no new information, since the existence of short-range magnetic correlations can already be inferred from the presence of a small but visible hump of diffuse scattering near where the first AF Bragg peak appears at lower temperature~\cite{shull;pr51,blech;p64,melle;jpcm98}. The unique contribution of mPDF analysis becomes more apparent with a close inspection of the mPDF fits above \TN, which reveal that the long-range Type~II AF structure does not actually fit the mPDF data well, particularly for the first few nearest neighbors. For example, the experimental mPDF peaks associated with the 2nd- and 4th-nearest neighbor spin pairs near 4.4~\AA\ and 6.3~\AA\ are consistently much higher in magnitude than predicted by the long-range structure, as seen for the 120~K fit in Fig.~\ref{fig:LRO}. This indicates that the short-range magnetic correlations in the paramagnetic (PM) phase deviate significantly from the long-range structure and cannot be viewed simply as the same type of magnetic order on a weaker or shorter length scale. These subtle differences can provide important insights into the underlying physics and may help differentiate competing theories.
	
	To pursue this further, we turned to the ``disordered local moment'' DLM-DFT of finite-temperature magnetism~\cite{gyorf;jppfm85} in which strong electronic correlations evident in transition metal oxides~\cite{hughe;njp08} and lanthanide compounds~\cite{hughe;n07,petit;prl15}, for example, are treated with the self-interaction correction (SIC)~\cite{luder;prb05}. When applied to MnO, the theory accounts for the magnitude of the insulating gap and its persistence into the paramagnetic phase, the magnetic ordering temperature and the onset of the type-II AF state. Our first-principles DLM-DFT calculations find the ground state Mn-ion configuration to be Mn$^{2+}$ with five localized d-states constituting a half-filled shell in line with Hund's first rule. 
	
	Without prior assumption of any specific effective spin Hamiltonian, the key quantity calculated by DLM-DFT is the lattice Fourier transform (LFT), ${\cal S}^{(2)}({\bf q})$, of the direct correlation function ${\cal S}^{(2)}_{0\,n}$ for the local Mn moments, from which the magnetic correlation function, $\langle\boldsymbol{S}_0 \cdot \boldsymbol{S}_n\rangle$, can be determined. Here, $\boldsymbol{S}_0$ represents a classical Heisenberg spin-like local moment arbitrarily chosen to be at the origin, $\boldsymbol{S}_n$ represents a spin in the $n$th coordination shell (assuming the ideal cubic structure), and the angled brackets denote an average over all the spins in that shell.  We find the $\langle\boldsymbol{S}_0 \cdot \boldsymbol{S}_n\rangle$'s using an Onsager cavity field calculation~\cite{staun;prl92} by solving the coupled integral equations, $C({\bf q})^{-1} = \left(1 - \beta ({\cal S}^{(2)}({\bf q}) - \Lambda)\right)$ and $\Lambda = \int {\cal S}^{(2)}({\bf q})\,C({\bf q})\, d{\bf q}$ for the LFT of the magnetic correlation function, $C({\bf q})$. This pair of equations ensures that the sum rule  $< S_0^2 > = 1$ is met.  Furthermore the real space ${\cal S}^{(2)}_{0\,n}$ quantities describe the magnetic exchange interactions, $J_n$'s, between the Mn spins on different shells. When analyzed in concert with the underlying electronic structure~\cite{hughe;njp08} an \abinitio\ prediction of a ``textbook'' picture of Anderson superexchange between O 2p and Mn 3d orbitals emerges.
	
	\begin{figure}
		\includegraphics[width=80mm]{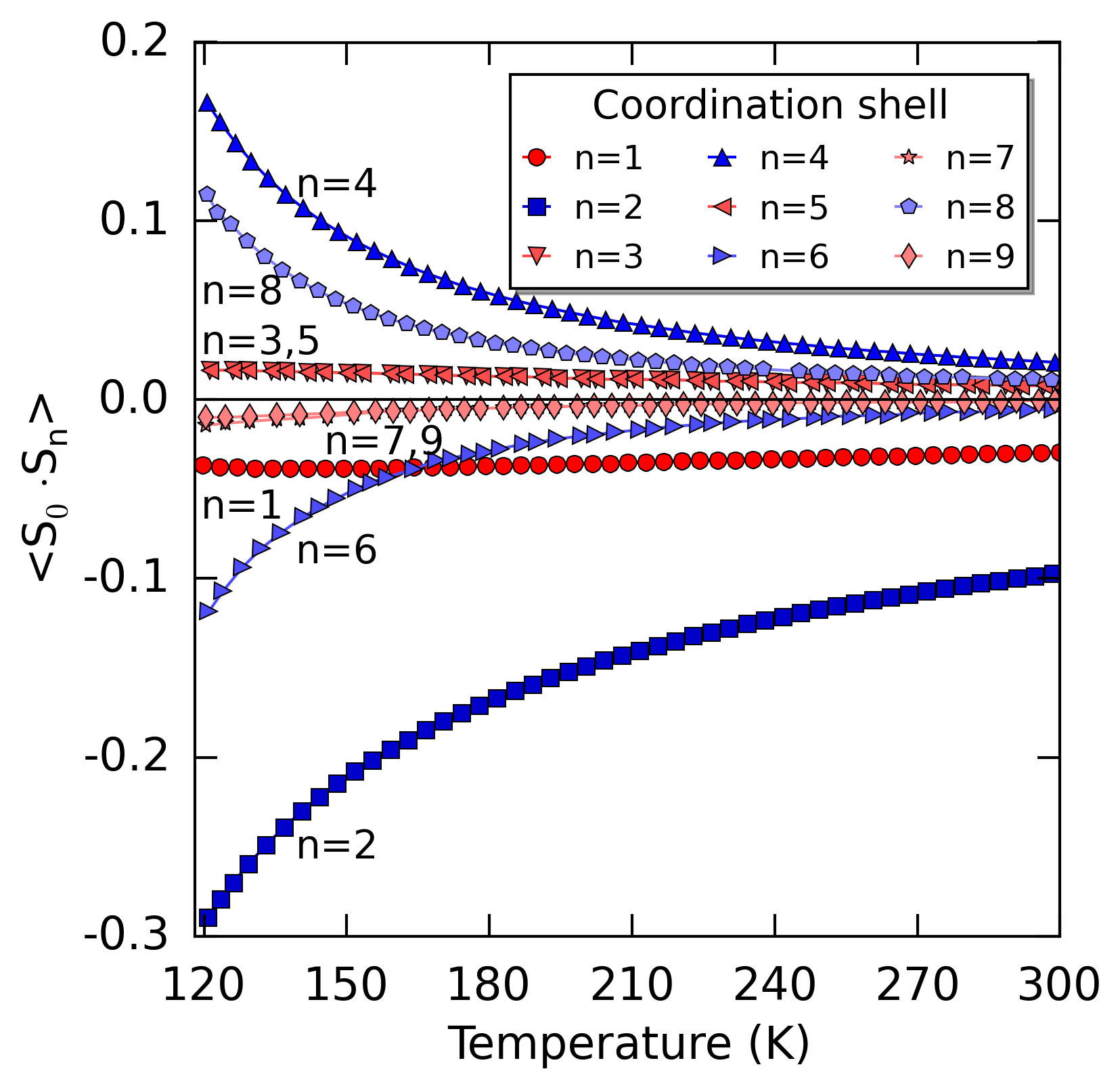}
		\caption{\label{fig:theory} (Color online) Temperature-dependent magnetic correlation functions of MnO for the first 9 nearest neighbors of the cubic structure calculated from the \abinitio\ theory described in the text. }
	\end{figure}
	
	In Fig.~\ref{fig:theory}, we display the magnetic correlation functions $\langle\boldsymbol{S}_0 \cdot \boldsymbol{S}_n\rangle$ calculated as a function of temperature from the \abinitio\ theory. The \abinitio\ calculations were done in terms of the dimensionless temperature ratio $T/T_{\mathrm{N}}$, which we converted to a physical temperature using the known \TN\ of 118~K. It is notable that the magnetic correlations with odd-numbered shells, shown in red hues, are significantly smaller (though more independent of temperature) than those with even-numbered shells (blue hues). This reflects the geometric frustration of the Type II AF structure on the face-centered-cubic Mn sublattice, in which half of the neighbors in the odd-numbered shells lie within the same (111) plane as the central spin $\boldsymbol{S}_0$ and would therefore tend to be parallel to it, while the other half are in planes with opposite magnetization in the ordered state. In contrast, the neighbors in even-numbered shells suffer no such frustration and can all have uniform tendency for either parallel or anti-parallel alignment with the central spin.

\begin{figure}
		\includegraphics[width=80mm]{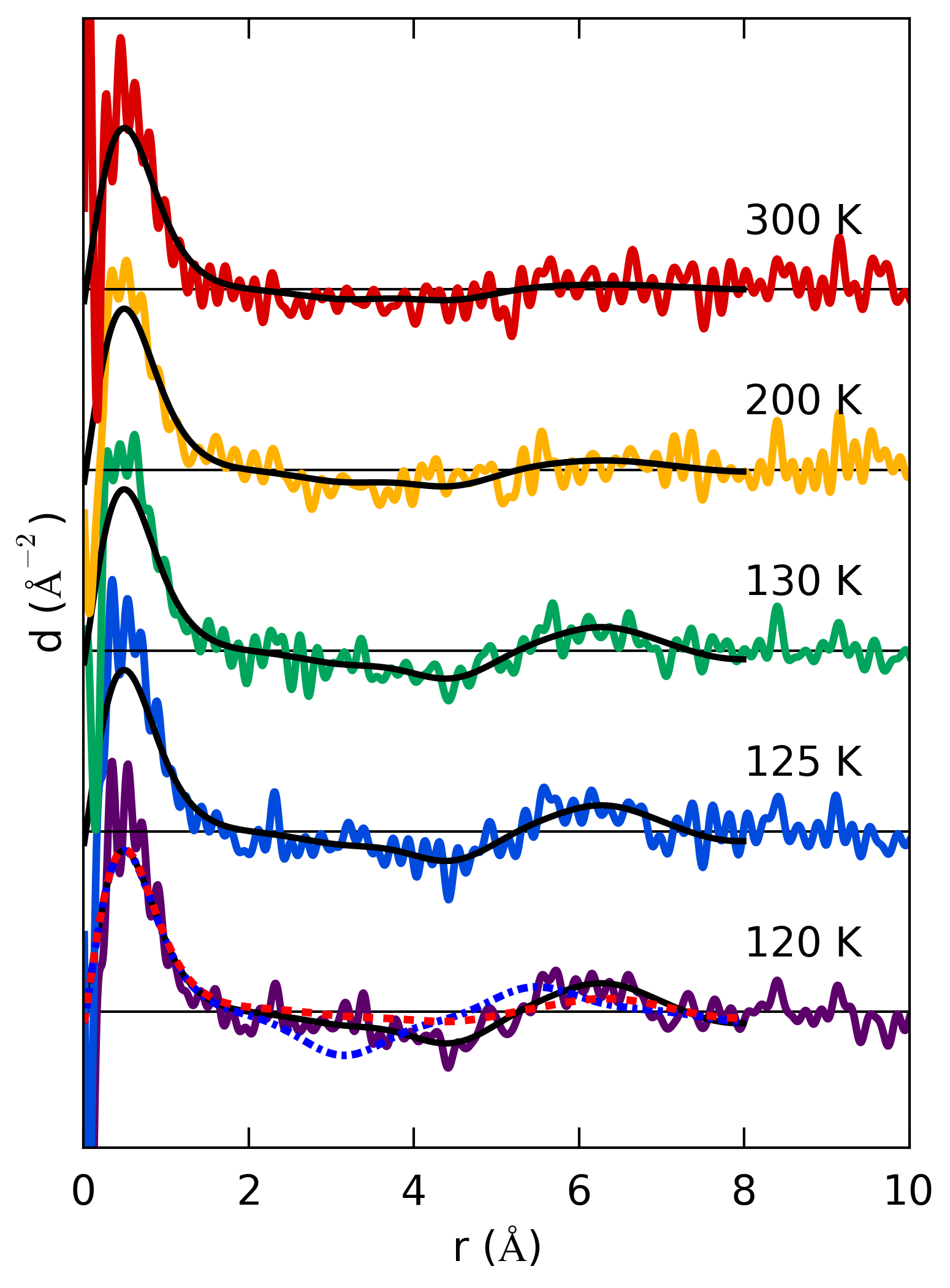}
		\caption{\label{fig:SRO} (Color online) Experimental (thick colored curves) and calculated (thinner black curves) mPDFs of MnO at 5 temperatures in the paramagnetic state. For 120~K, the best fit of the low-temperature long-range-ordered structure is shown as a dashed red curve, and the calculated mPDF for a larger $J_1/J_2$ ratio simulating a direct exchange scenario is shown as the blue dashed-dotted line.}
	\end{figure}

	To compare the theory to experiment, we calculated the individual mPDF contributions from a cubic model of the first 9 near-neighbor coordination shells determined by atomic PDF fits and scaled them according to the predicted correlation function at the appropriate temperature. The results of this procedure are displayed for the 5 measurements obtained in the paramagnetic state in Fig.~\ref{fig:SRO}.
	The calculated mPDF functions are shown as black solid curves overlaid on the data. The agreement between theory and experiment is excellent at all temperatures, especially considering that there are no free parameters in these calculations. For comparison, the best fit of the long-range magnetic structure is shown for 120~K as the red dashed curve. It clearly describes the data much worse than the \abinitio\ prediction, particularly for the 2nd- and 4th-nearest neighbors at $\sim$4.4~\AA\ and $\sim$6.3~\AA. These experimental results are in general agreement with earlier magnetic reverse Monte Carlo analysis of MnO~\cite{melle;jpcm98}. Further details regarding the structural and magnetic PDF refinements and calculations are found in the Supplementary Information. 
	
	The mPDF data allow us to compare different \abinitio\ theories. A common difference among \abinitio\ theories is the first- and second-nearest-neighbor exchange parameter ratio $J_1 / J_2$, which can vary widely from more than 3.5 to less than 0.4 (see e.g. Table III in Ref.~\cite{fisch;prb09}). The present DLM-DFT method, producing essentially Anderson superexchange, predicts $J_1/J_2=0.502$. Higher ratios, such as 3.571 given by recent tight-binding calculations~\cite{harri;prb08}, are typically reflective of a substantial Mn-Mn direct exchange contribution in addition to the superexchange. The calculated temperature-dependent correlation functions can easily be modified to reflect different $J_1/J_2$~ ratios, from which the mPDF can be generated and compared to experiment. We show the result for $J_1/J_2=2.0$ at 120~K as the blue dashed-dotted curve in Fig.~\ref{fig:SRO}. Unsurprisingly, the larger ratio leads to stronger nearest-neighbor correlations, causing the calculated mPDF to differ significantly from the data. The fits suggest that the $J_1/J_2$ ratio is smaller, and close to 0.5, which supports a picture of there being Anderson superexchange without any significant contribution from direct exchange.
	
	To be more quantitative, we calculated the mPDF at the five temperatures above \TN\ for 17 different $J_1/J_2$~ratios distributed on the interval [0,1.2]. For each combination of temperature and exchange ratio, we calculated the summed squared difference between data and calculation ($\chi^2$). We then fit a parabola to the 17 ($J_1/J_2$, $\chi^2$) ordered pairs for each temperature, with a common parabolic minimum refined as a global parameter to obtain the best-fit value of $J_1/J_2$. We took care to properly weight with the statistical uncertainties on the data propagated from the original PDF on the Nyquist grid~\cite{farro;prb11}. Details are available in the Supplementary Information. This procedure revealed an optimal exchange ratio of $J_1/J_2=0.67\pm0.12$, in fairly close agreement with the predicted value from DLM-DFT and well away from larger values indicative of significant direct exchange contributions. This increases our confidence in the superexchange-dominated picture obtained from the DLM-DFT approach. Although experimental estimates of $J_1/J_2$ are few in number and vary from significantly less than 1 to well over 2, most place $J_1/J_2$ in the range of 0.8-0.9, in relatively good agreement with the present results~\cite{lines;pr65,kohgi;ssc72,bloch;prb73,pepy;jpcs74}.
	
	These findings constitute experimental and theoretical verification of the Anderson superexchange model of magnetic interactions in MnO. The success of the DLM-DFT \abinitio\ scheme for MnO suggests that it should be a good starting point for other TMOs, and coupled with measurements such as mPDF, will be of significant utility for understanding more complex materials. We have also demonstrated that the mPDF method can yield unique insights into magnetic systems by allowing quantitative analysis of short-range magnetic correlations directly in real space. We expect this to be valuable in the study of strongly correlated systems and other materials known to have short-range magnetism.

	\begin{acknowledgments}
		We thank Joan Siewenie for technical assistance with the measurements performed on the NPDF instrument. BAF and YJU acknowledge support from the NSF via PIRE program
		OISE-0968226 and DMREF program DMR-1436095, and BAF by the NSF GRFP
		program DGE-11-44155. SJLB acknowledges support from the U.S. Department of Energy, Office of Science, Office of Basic Energy Sciences (DOE-BES) under contract No. DE-SC00112704. JBS acknowledges support from the EPSRC (UK) grant EP/J006750/1. Neutron scattering experiments were carried out on NPDF at the Lujan Center, funded by DOE Office of Basic Energy Sciences. LANL is operated by Los Alamos National Security LLC under DOE Contract No. DE-AC52-06NA25396.
	\end{acknowledgments}

\end{document}


	
	\title{Supplementary Information: Verification of Anderson superexchange in MnO via magnetic pair distribution function analysis and \textit{ab initio} theory}
	
	
	\author{Benjamin A. Frandsen}
	\affiliation{Department of Physics, Columbia University, New York, NY 10027}
	
	\author{Michela Brunelli}
	\affiliation{Swiss Norwegian Beamlines, European Synchrotron Radiation Facility (ESRF), Grenoble, France}
	
	\author{Katharine Page}
	\affiliation{Spallation Neutron Source, Oak Ridge National Laboratory}
	
	\author{Yasutomo J. Uemura}
	\affiliation{Department of Physics, Columbia University, New York, NY 10027}
	
	\author{Julie B. Staunton}
	\affiliation{Department of Physics, University of Warwick, Coventry CV4 7AL, United Kingdom}

	\author{Simon J. L. Billinge}
	\email{sb2896@columbia.edu}
	\affiliation{%
		Condensed Matter Physics and Materials Science Department, Brookhaven National Laboratory, Upton, NY 11973, USA.
	}%
	\affiliation{%
		Department of Applied Physics and Applied Mathematics, Columbia University, New York, NY 10027, USA.
	}%
	
	\date{\today}
	
	
	\maketitle
	
	\section{Structural and magnetic PDF refinements}
\begin{figure}
		\includegraphics[width=60mm]{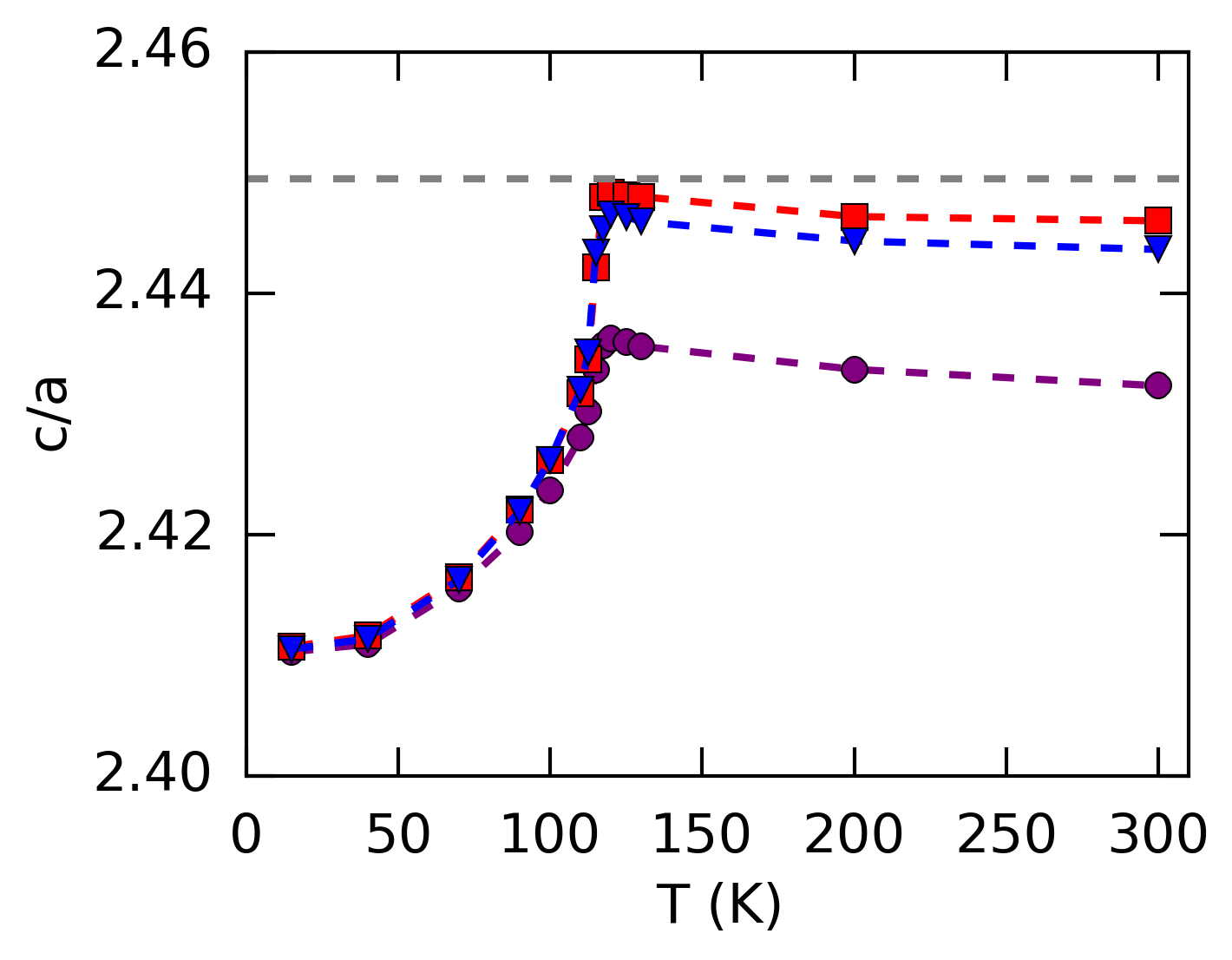}
		\caption{\label{fig:latparams} Ratio of the refined rhombohedral lattice parameters $a$ and $c$ obtained from fits over $r$-ranges of 0-20~\AA\ (purple circles), 0-50~\AA\ (blue triangles), and 0-100~\AA\ (red squares). The gray dashed horizontal line at $\sqrt{6}\simeq 2.449$ represents the ratio corresponding to the ideal cubic structure.}
	\end{figure}

	PDFs were generated from PDFgetN~\cite{peter;jac00} using a maximum momentum transfer of $Q_{\mathrm{max}}$=35~\AA$^{-1}$. The PDF refinements were carried out using the $R\overline{3}m$ rhombohedral structural model at all temperatures. Although a slight monoclinic distortion was previously found at low temperatures~\cite{goodw;prl06,frand;aca15}, the associated changes to the atomic positions are small enough to have very little effect on the mPDF and have therefore been neglected in the present analysis. Interestingly, the atomic structure remains slightly rhombohedral on a local length scale in the high-temperature paramagnetic phase. Supp. Fig.~\ref{fig:latparams} displays the lattice parameter ratio $c/a$ as a function of temperature for fits carried out to 20~\AA\ (purple circles), 50~\AA\ (blue triangles), and 100~\AA\ (red squares). The gray horizontal dashed line represents the ratio corresponding to perfect cubic symmetry in the rhombohedral basis. It is clear that the ratio in the paramagnetic temperature region approaches the ideal value as the $r$-range of the fit is increased, indicating that the short-range structure remains slightly rhombohedral while the long-range average structure is cubic, as has been established from standard diffraction measurements. The persistence of the local rhombohedral structure is likely to be closely related to the persistence of short-range magnetic correlations.
	
	After performing the structural PDF fit, the calculated PDF was subtracted from the data, leaving the experimental mPDF and any residual background signal or errors from the structural refinement. We also normalized by the refined structural PDF scale factor to allow for meaningful comparison of the magnitude of the mPDF at various temperatures. We observed that a low-amplitude, temperature-independent, sinusoidal background signal was present in the PDF at all temperatures, arising from sharp edges in the neutron scattering data most likely due to imperfect shielding of the instrument. To ensure that it did not confound the mPDF fits, we isolated this background signal by subtracting out the best-fit PDF+mPDF calculation at 15~K, and we then subtracted it from all other data sets. For all temperatures except 15~K, the PDF data shown in the paper have this background removed.
	
	We then fit the long-range-ordered Type II antiferromagnetic structure to these scale-normalized, background-corrected mPDF curves, using Eq.~1 in the main text. The fits were performed within the Python programming language using least-squares minimization routines contained in the NumPy and SciPy libraries. For simplicity, we fixed the spin directions to lie within the (111) plane, as is well known from earlier diffraction experiments and confirmed by our earlier mPDF study~\cite{frand;aca15}. Initially, we allowed both scale factors in Eq.~1, $C_1$ and $C_2,$ to vary independently. We found that $C_2$, the scale factor corresponding to the structureless part of the unnormalized mPDF, had little variation with temperature (as expected when the local moment size is constant), so we fixed it to the average of the refined values for each temperature, leaving $C_1$ as the only adjustable parameter in the fits. As discussed in the main text, these simple fits described the data taken below \TN\ very well, but were unable to accurately capture the short-range correlations above \TN.
	
	\section{Calculation of the magnetic PDF in the paramagnetic phase}
	The DLM-DFT \textit{ab initio} calculations produced the temperature-dependent magnetic correlation functions $\langle\boldsymbol{S}_0 \cdot \boldsymbol{S}_n\rangle$ shown in Fig.~3 of the main text. As explained in the main text, these quantities are averaged over all the spins in the $n$th coordination shell of the ideal cubic structure. The theory predicts only the dot product of the spin pairs, not the absolute orientations of each spin. The calculated mPDF has no sensitivity to the absolute spin orientation in the ideal cubic structure, so an arbitrary direction was chosen as the spin orientation axis. Ignoring the slight rhombohedral distortion of the lattice, the mPDF contribution from each cubic coordination shell was first calculated assuming perfect alignment of the spins, i.e. with a scale factor equal to $-C_2 / \sqrt{2\pi}$. This scale factor was then multiplied by the absolute value of $\langle\boldsymbol{S}_0 \cdot \boldsymbol{S}_n\rangle$ for that particular shell at the appropriate temperature. The individual contributions from the first 9 coordination shells were then added together and overlaid on the data, producing Fig.~4 in the main text.
	
	\section{Determination of optimal $J_1 / J_2$ ratio}
	To determine the optimal $J_1/J_2$ ratio, we generated spin correlation functions and the corresponding mPDFs at the five measured temperatures above \TN\ for 17 values of $J_1/J_2$ distributed approximately uniformly on [0,1.2]. To ensure that the statistical uncertainties on the original data were properly handled, we generated the experimental PDFs with their associated statistical uncertainties on the Nyquist grid~\cite{farro;prb11} and propagated these uncertainties through each step of the analysis outlined below. The procedure to determine the optimal $J_1/J_2$ ratio is as follows:
	\begin{enumerate}
		\item  Extract the mPDF at each temperature by refining the structural model on the Nyquist grid and subtracting the calculated best-fit PDF. Normalize by the scale factor and subtract the temperature-independent background as described previously.
		\item For each temperature, generate the mPDF for each of the 17 $J_1 / J_2$ values.
		\item Compute the summed squared difference $\chi^2$ between the experimental calculated mPDF for each $J_1 / J_2$ value at each temperature. Explicitly, this is given by $\chi^2(T, J_1/J_2) = \sum (y_{\mathrm{obs}}-y_{\mathrm{calc}})^2$, where $y_{\mathrm{obs}}$ and $y_{\mathrm{calc}}$ are the observed and calculated mPDFs, $T$ and $J_1/J_2$ are the corresponding temperature and exchange parameter ratio, and the sum is taken over all data points on the range $0.09\le r \le 8.0$~\AA. This yields 17 ($J_1/J_2,\chi^2$) ordered pairs for each temperature. Each ordered pair has an associated statistical uncertainty propagated from the original PDF data set on the Nyquist grid.
		\item For each temperature, use the Levenberg-Marquardt least-squares minimization algorithm to fit a parabola of the form $y = A(x-x_0)^2+C$ to the corresponding ($J_1/J_2,\chi^2$) ordered pairs, with weights given by the statistical uncertainty on each $\chi^2$ point. The parameters $A$ and $C$ are independent for each temperature, but $x_0$ is globally refined for all 5 temperatures, ensuring a common minimum. 
	\end{enumerate}
	
	This procedure resulted in an optimal value of $x_0 = J_1/J_2=0.67\pm0.12$, as reported in the main text.

	\section{Assessing the robustness of the mPDF signal}
\begin{figure}
		\includegraphics[width=80mm]{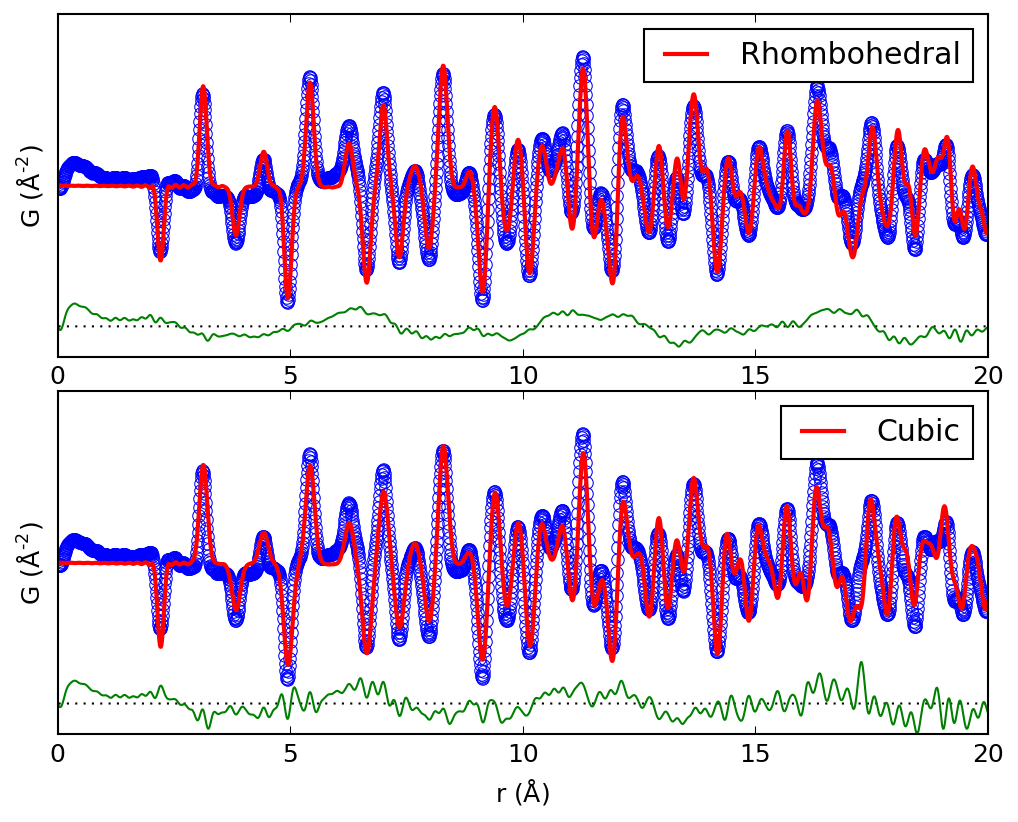}
		\caption{\label{fig:resids} Atomic PDF fits to the MnO total PDF at 15~K, with the data shown as blue circles, the calculated model as the red curve, and the offset difference as the green curve. Top panel: Refinement using the rhombohedral structural model. Bottom panel: Refinement using the cubic model, which is known to be incorrect below \TN\ = 118~K.}
	\end{figure}

	Since the experimental mPDF is extracted from the fit residual of the refined atomic PDF model against the total PDF, it is important to assess the risk of any imperfections in the atomic PDF model interfering with the experimental mPDF. As a rigorous consequence of the fact that the magnetic scattering is not normalized by the magnetic form factor when the total PDF is generated, the mPDF signal is broadened out significantly in real space~\cite{frand;aca15}. Additionally, since only the Mn ions contribute magnetic moments, the spatial frequencies of the mPDF signal are significantly lower than for the atomic PDF, to which both Mn and O contribute. Therefore, the mPDF signal is easily distinguished from the atomic PDF and, importantly, from features that propagate to the atomic PDF difference curve from an imperfect structural fit: the mPDF signal is broad and slowly varying, while contributions to the atomic PDF fit residual from an imperfect model are much higher in frequency. To illustrate the differences between the mPDF and typical imperfections in the atomic PDF fit, we display in Supp. Fig.~\ref{fig:resids} two atomic PDF fits to the 15 K data set, the first one using the established rhombohedral structure, and the second one using the incorrect cubic structure.

	In both cases, the broad, long-wavelength oscillations of the mPDF are clearly evident in the green fit residual curve; however, the cubic fit also has a high-frequency component in the fit residual coming from the inability of the cubic structure to perfectly describe the structural PDF data. 

	To further establish the robustness of the mPDF signal against different factors in the data processing, we repeated the procedure for obtaining the mPDF for several different combinations of Fourier transform settings to obtain the total experimental PDF. We selected three different values of $Q_{\mathrm{max}}$, i.e. the maximum momentum-transfer included when computing the Fourier transform of the total scattering data, and for each value of $Q_{\mathrm{max}}$, we generated the experimental PDF on two different grids in real space: a fine grid with a spacing of 0.01~\AA, and a much coarser grid with a spacing of $\pi$/$Q_{\mathrm{max}}$ representing the Nyquist sampling grid. 
	\begin{figure}
		\includegraphics[width=80mm]{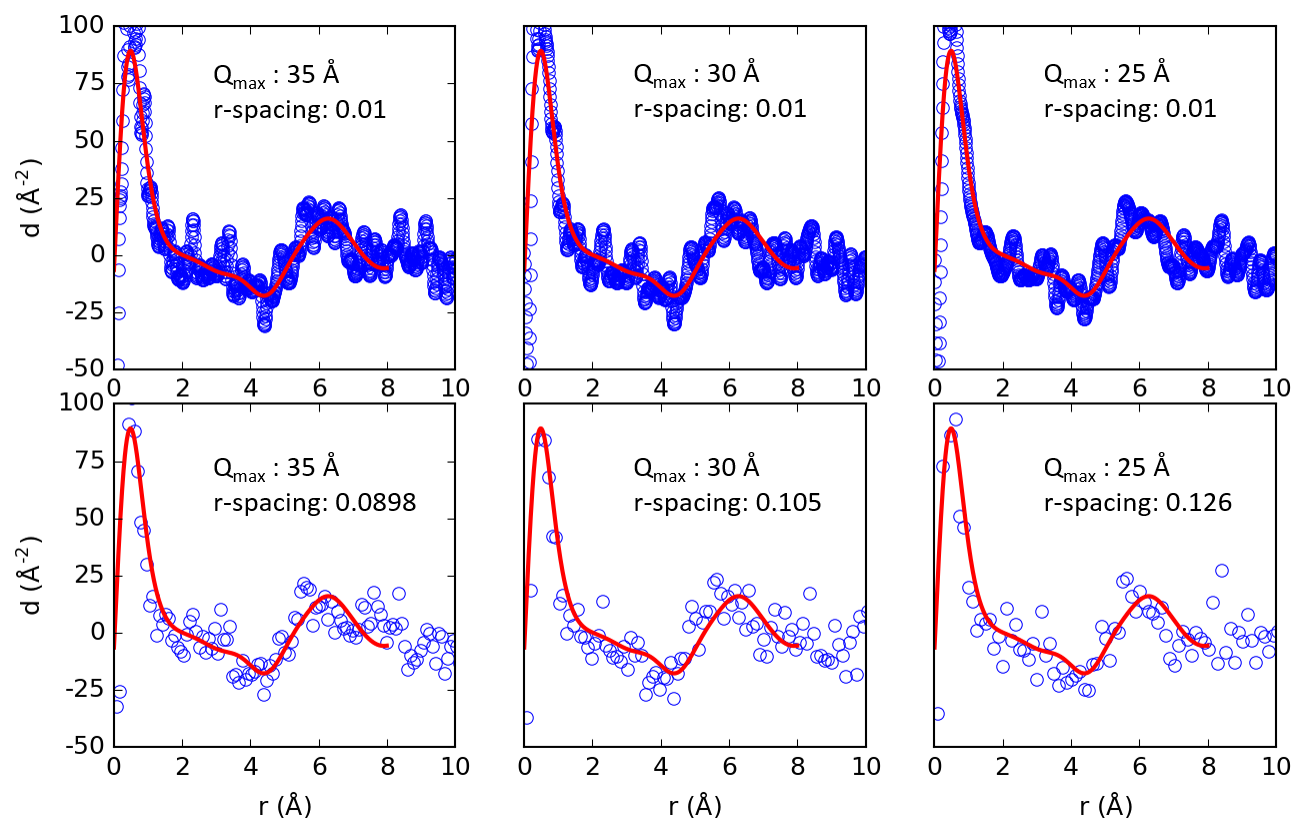}
		\caption{\label{fig:transforms} Experimental mPDF at 120~K (blue circles) for various Fourier transform parameters indicated in each panel, with the \abinitio\ calculated mPDF overlaid as the red curve for comparison.}
	\end{figure}
	Artifacts from the data reduction process that could possibly interfere with the mPDF signal would be expected to be different for these various Fourier transform protocols, so if the putative mPDF signal were to remain consistent across all of them, it would greatly boost our confidence in the reliability of the data. This was, in fact, the case, as illustrated in Supp. Fig.~\ref{fig:transforms} showing the mPDF signal at 120~K obtained from subtraction of the atomic PDF for each of the Fourier transform combinations, with the \abinitio\ calculated mPDF overlaid for comparison. The agreement is excellent in all cases, providing strong support for the reliability of this procedure for extracting the mPDF.